\documentclass[10pt,oneside]{article}
\usepackage{geometry}
\usepackage{graphicx}
\usepackage{graphics}
\usepackage{amsfonts}
\usepackage{amsmath}
\usepackage[english]{babel} 
\usepackage{xcolor}
\begin{document}
\title{Cascading traffic jamming in a two-dimensional Motter and Lai model.}
\author{Gabriel Cwilich$^1$, and Sergey V. Buldyrev$^1$\\
$^1$ Department of Physics, Yeshiva University, 500 West 185th Street, New
York, New York 10033\\
}
\begin{abstract}
  We study the cascading traffic jamming on a two-dimensional random
  geometric graph using the Motter and Lai model. The traffic jam is
  caused by a localized attack incapacitating a circular region or a
  line of a certain size, as well as a dispersed attack on an equal
  number of randomly selected nodes. We investigate if there is a
  critical size of the attack above which the network becomes
  completely jammed due to cascading jamming, and how this critical size
  depends on the average degree $\langle k\rangle$ of the graph,
  on the number of nodes $N$ in the system, and the tolerance parameter $\alpha$ of the
  Motter and Lai model. \end{abstract} \date{\today}
\maketitle

\section{Introduction}
The Motter and Lai model~\cite{motterLai,motter} is a generic model of
overload spreading. It can be used to model street traffic~\cite{Havlin,Perez}, internet
traffic~\cite{motterLai} and power grids~\cite{Kornbluth2018,Kornbluth2021,yang,Anghel}.  In the Motter
and Lai model, the betweenness of a node is defined as the number of
the shortest paths connecting any pair of nodes in the network that
pass through (but do not end in) this node. After the network is
constructed, the initial betweenness $b^{(o)}_i$ of each node $i$ is
calculated. A node can withstand a maximum betweenness of
$L_i\equiv (1+\alpha)b^{(o)}_i$, where $\alpha$, the tolerance, is a
global parameter of the system. When modeling traffic, a node should be
understood as an intersection of several streets or roads, which
becomes gridlocked if the amount of cars trying to cross it in a unit
time exceeds a limit set by the tolerance. Previously, cascading
failures in the Motter and Lai Model were studied in randomly connected
networks not embedded in space~\cite{Kornbluth2018,Kornbluth2021},  and
on a square lattice where the shortest path was defined by minimizing
the sum of the normally distributed weights of the links,  which may be
interpreted as the traveling times between  pairs of sites~\cite{Havlin}.  In the latter work a square of a given size is deleted
from the lattice. As a result, the new shortest paths trying to avoid
the damaged area concentrate in the perimeter of the deleted square
and,  depending on the tolerance, the size of the square and the size
of the lattice, the nodes on the perimeter may become overloaded and
can be deleted from the system as unusable. As a result, at the next
stage of the cascade the nodes in the vicinity of the unusable nodes
may become overloaded and the traffic jam will spread around the
initially damaged area roughly as circles whose radii grow
approximately linearly with time, defined as the number of stages in the
cascade. The speed of growth increases as the tolerance $\alpha$ decreases. Also the
speed increases linearly with the size of the system. Among the important questions
not addressed so far, are the following. What is the minimal radius of the initially
damaged area for which a cascade starts to propagate? What is the
distribution of this minimal radius for different central nodes, and
how does it depend on the tolerance $\alpha$ and the size of the system? In
this paper we will address these questions for more realistic graphs than
the square lattice, namely the random geometric graphs~\cite{Bollobas}.
  
\section{The model}
We create a random geometric graph~\cite{Bollobas} on a unit square with periodic
boundaries by placing $N$ nodes with random coordinates $(x,y)$, each one 
uniformly distributed in the interval $(0,1)$. Next we select a radius
$r =\sqrt{\langle k\rangle/(\pi N)}$, where $\langle k\rangle$ is the
desired average degree of the graph, and connect all pairs of nodes
whose mutual distance is less or equal than $r$. By construction, this
graph has a Poissonian degree distribution with average degree
$\langle k \rangle$.  The percolation threshold for this graph is
$\langle k \rangle=k_c=4.512$~\cite{Bollobas}.  We use 
values of $\langle k\rangle$ larger than or equal to $5$ to ensure that the majority of
nodes belong to the giant component of the graph. Since only the giant
component  plays a role in the overall traffic on the
graph, we delete all the nodes in the finite clusters before unleashing
the initial attack on the system. We define the length of a path as
the sum of the Euclidean lengths of all its links. We calculate the initial betweenness of
each node $i$ of the graph,  $b_i$,  and set its maximal load as
$L_i=b_i(1+\alpha)$.

We produce three types of attacks on the graph (Fig.~\ref{fig1}): i) a
circular attack in which we delete the $n_d$ nodes that  have the shortest
path lengths to a randomly selected node called the center of the
attack; ii) a linear attack, in which we delete $n_d$ nodes selected from a
vertical stripe of width $r$ around the center of the attack,  in 
such a way that the standard deviation  in the
vertical direction from the center of the attack of the deleted nodes is minimal; and iii)
a random attack in which we delete $n_d$ nodes selected at random throughout the system.  

At stage $t=1$, immediately after the
initial attack, the betweenness of all nodes is recomputed and all the
nodes whose new betweenness exceeds $L_i$ are deleted. This process is
repeated for $t>1$ until the betweenness of any remaining node $i$
does not exceed $L_i$. For each realization of the network and of the initial
attack we compute the total number of cascading stages $N_t$, the
number of survived nodes at the end of the cascade $N_f$,  and the
number of nodes in the largest connected cluster at the end of the cascade $N_c$.  In addition,
for each stage $t$ of the cascade we compute the number of nodes
deleted at that stage, $F(t)$, and also their average shortest path distance $r(t)$
from the central node of the attack,  for the case of a circular or linear attacks.
For random attack this last  quantity is not defined.
\bigskip
\section{Results}
\subsection{Circular attacks} 
\begin{figure}[h!]
\centering
{
(a)
\includegraphics[width=.35\textwidth]{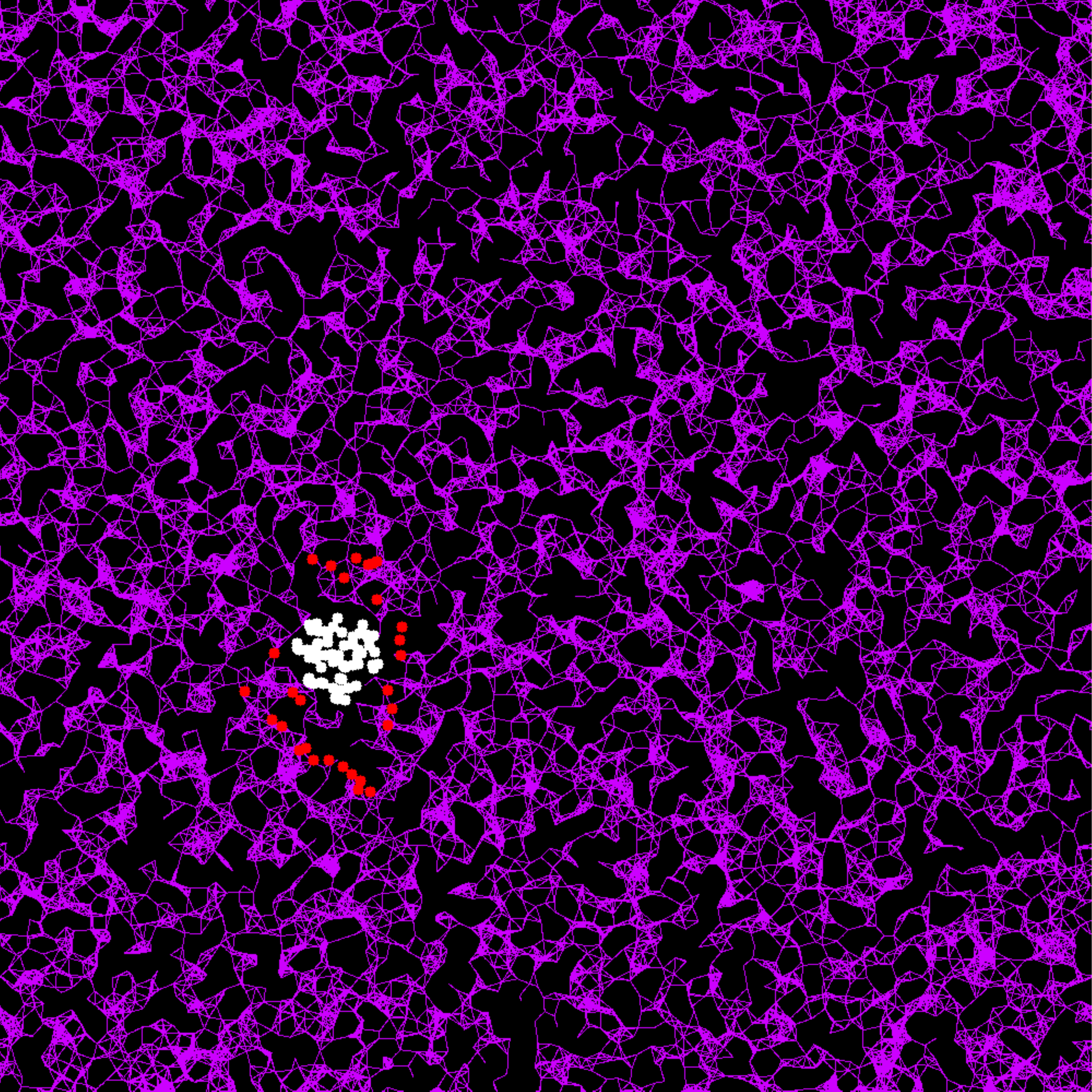}
(b)
\includegraphics[width=.35\textwidth]{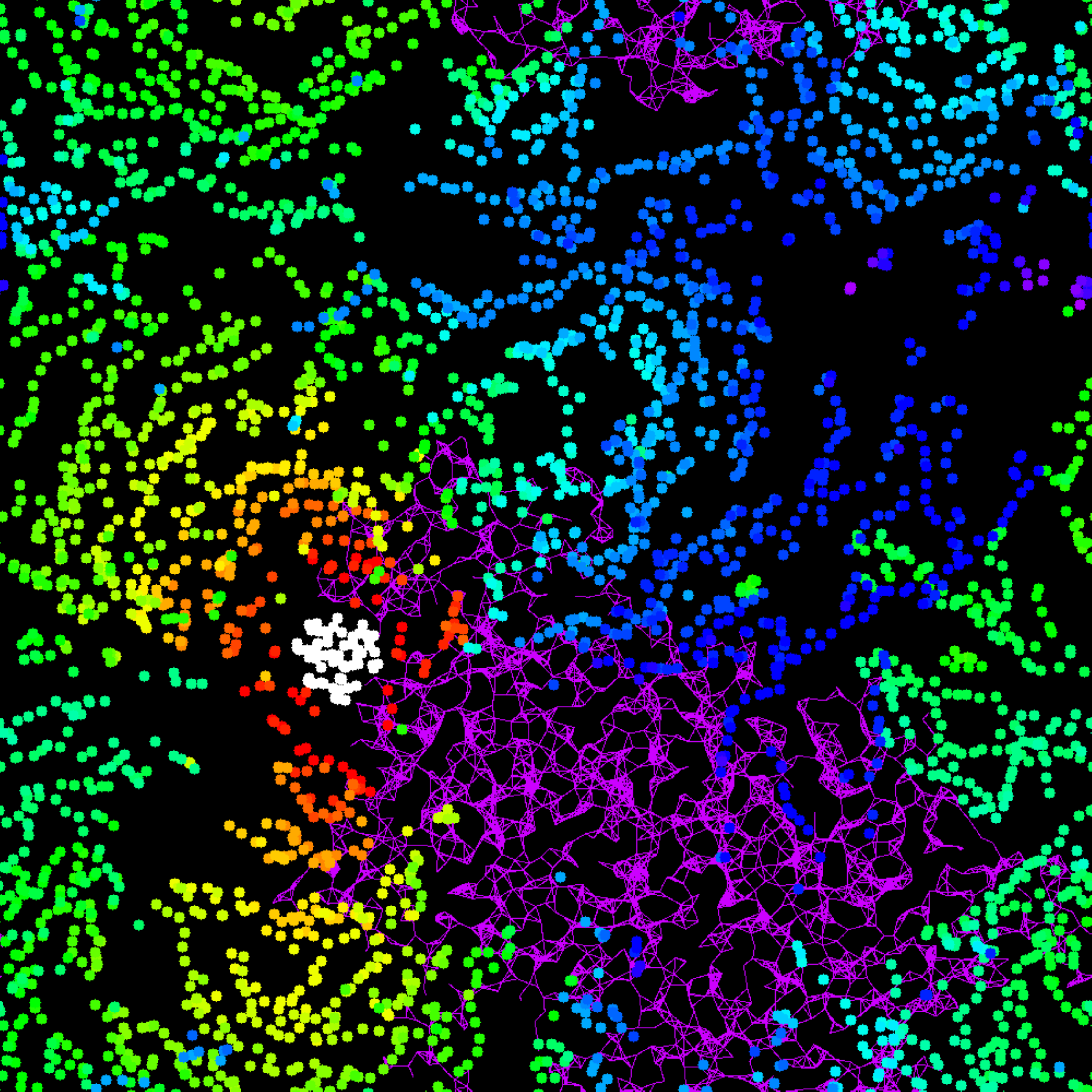}

(c)
\includegraphics[width=.35\textwidth]{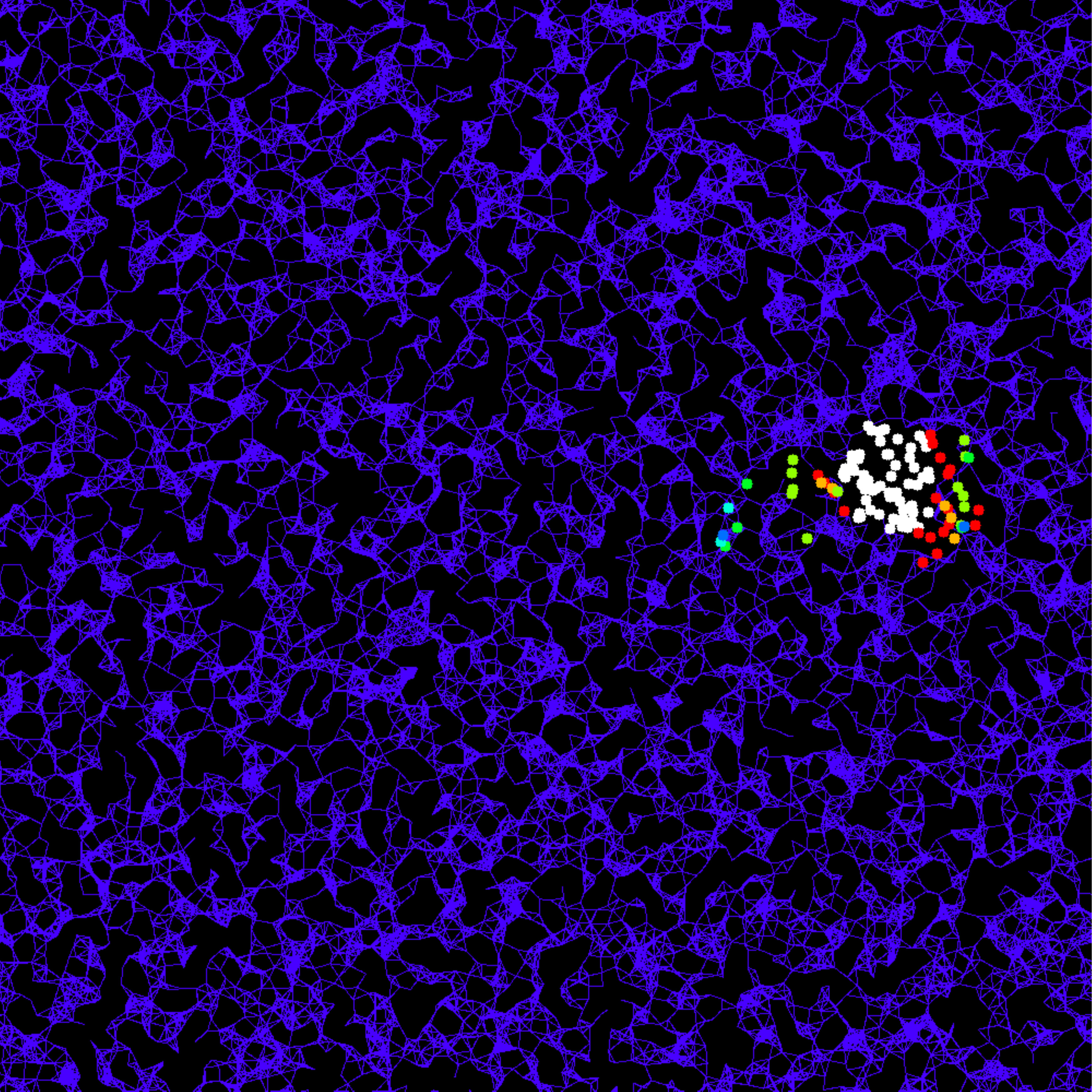}
(d)
\includegraphics[width=.35\textwidth]{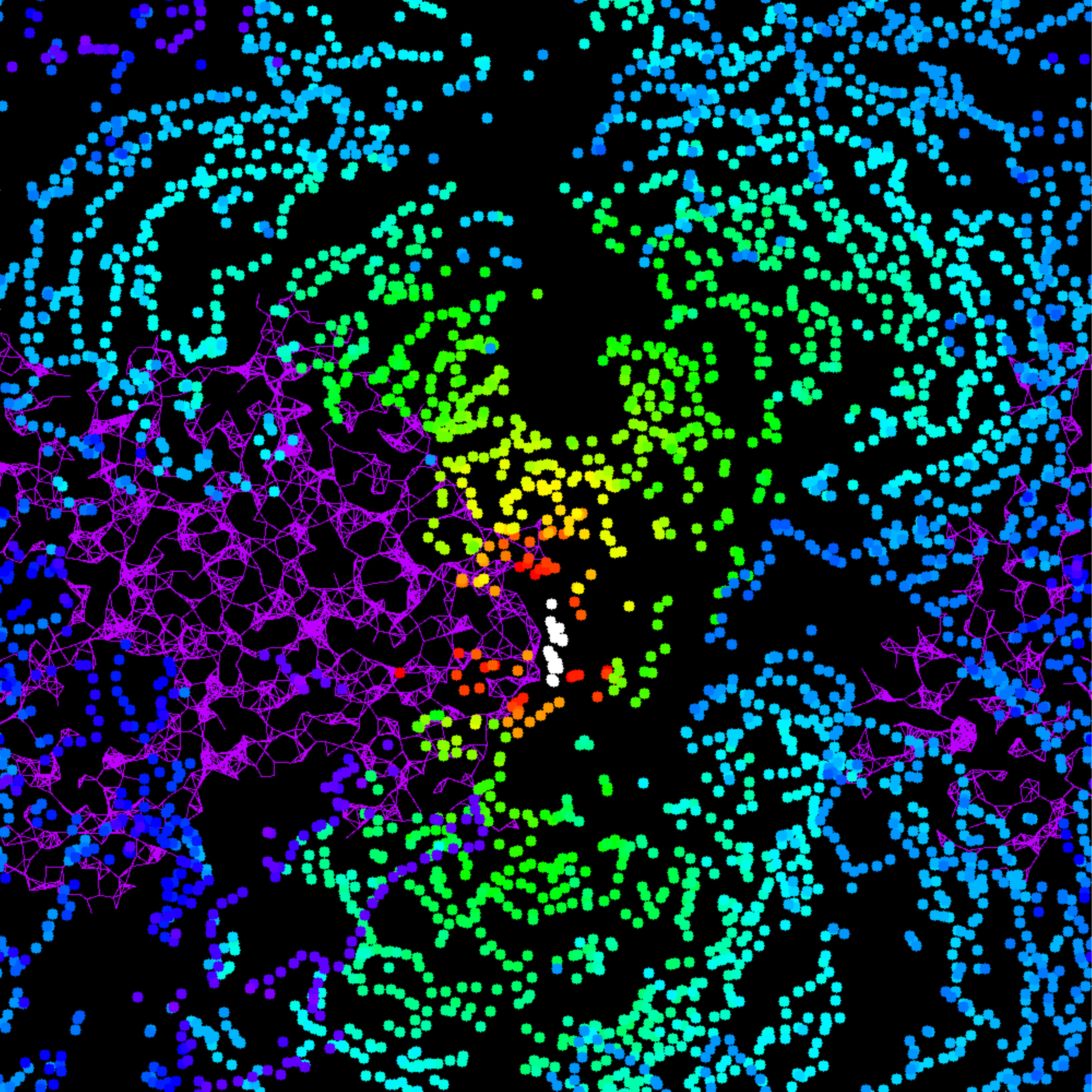}

(e)
\includegraphics[width=.35\textwidth]{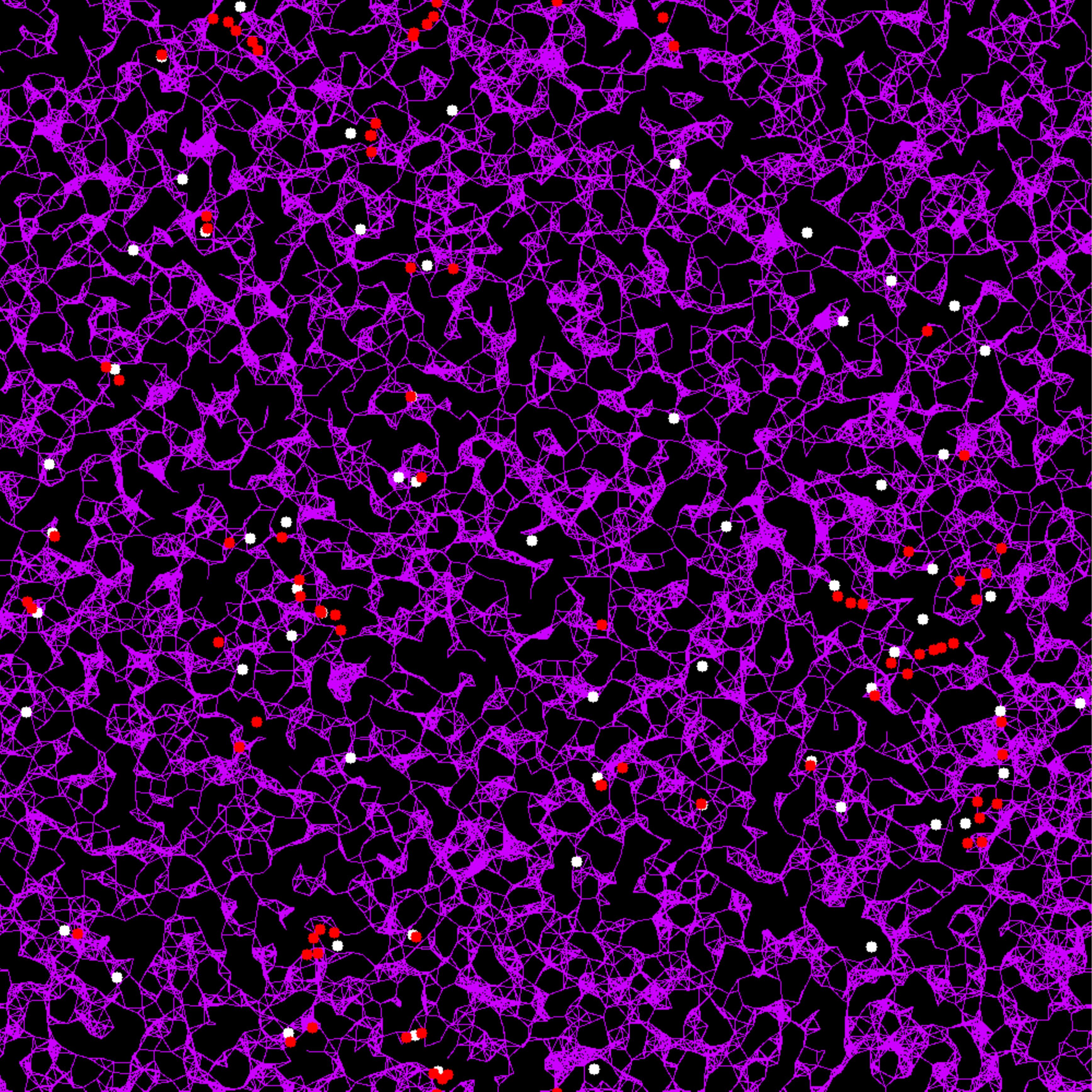}
(f)
\includegraphics[width=.35\textwidth]{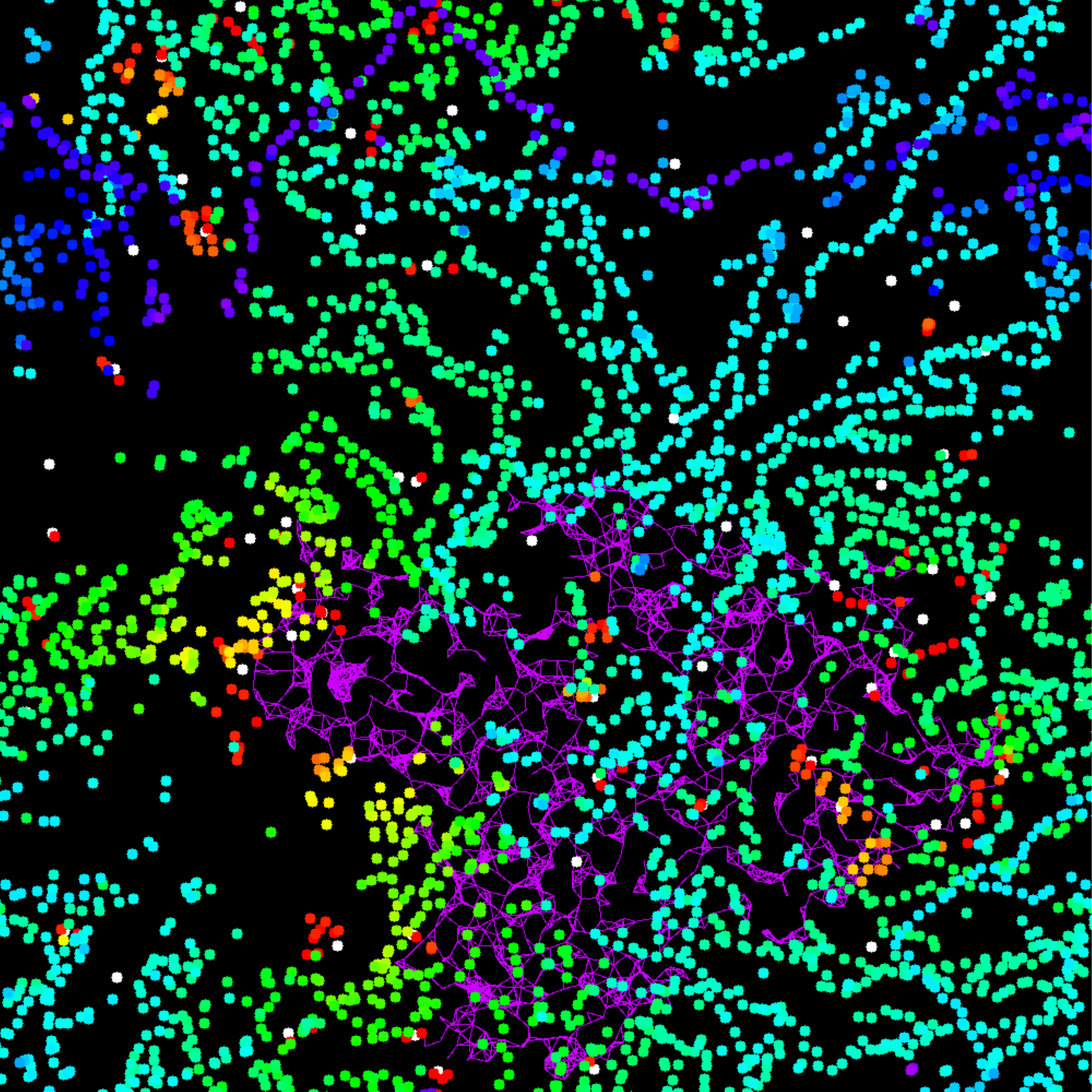}
}

\caption{
  \footnotesize (a) Circular attack on a geometric graph with $N=10000$, $\langle k\rangle=8$ and $\alpha=2$. The white circles represent the nodes attacked initially  ($n_d=64$). The red circles are the nodes overloaded at the first stage of the cascade.(b) The end of the cascade of the attack shown in (a). Different colors in rainbow order (red, orange, yellow, green, cyan, blue, violet) show the stages of the cascade from 1 to 37. 
  (c) Circular attack on $n_d=64$ nodes in a different graph, also  of $\langle k \rangle=8$ which ends after 7 stages without damaging the giant component.
  (d) Linear attack on the same graph with $n_d=16$ at the end of the cascade at stage 40. The color code is the same as in panel (b). (e) Stage 1 of the cascade after a random attack with $n_d=64$. (f) Final stage 37 of this random attack. The color code is the same as in panel (b). One can see that the overloads on the first stage are spread uniformly but at later stages (yellow and green) they are concentrated near three centers, resembling the patterns in panels (b) and (d).}
\label{fig1}
\end{figure}

\begin{figure}[h!]
\centering
{
  \includegraphics[width=.49\textwidth]{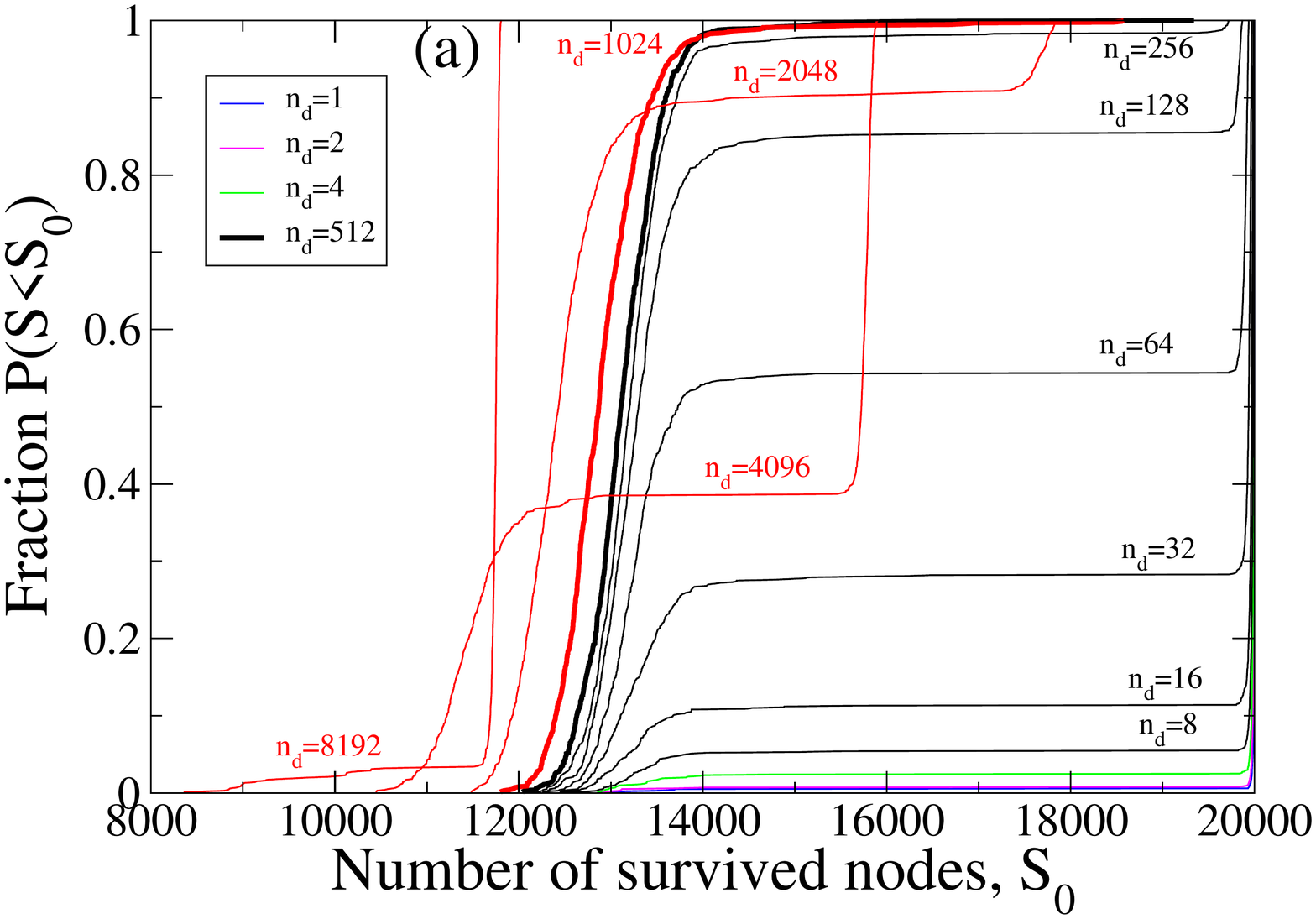}
  \includegraphics[width=.49\textwidth]{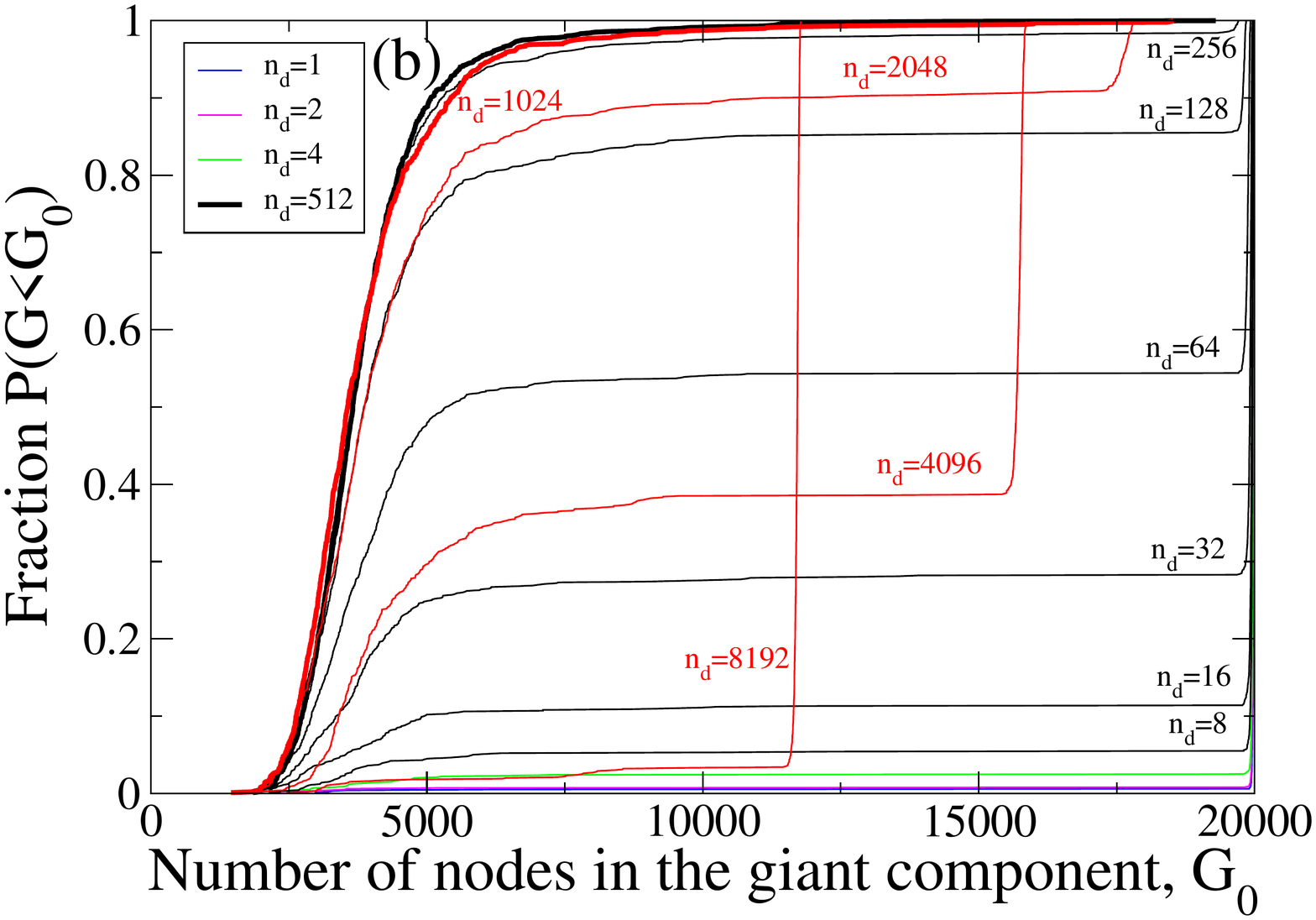}
  \includegraphics[width=.49\textwidth]{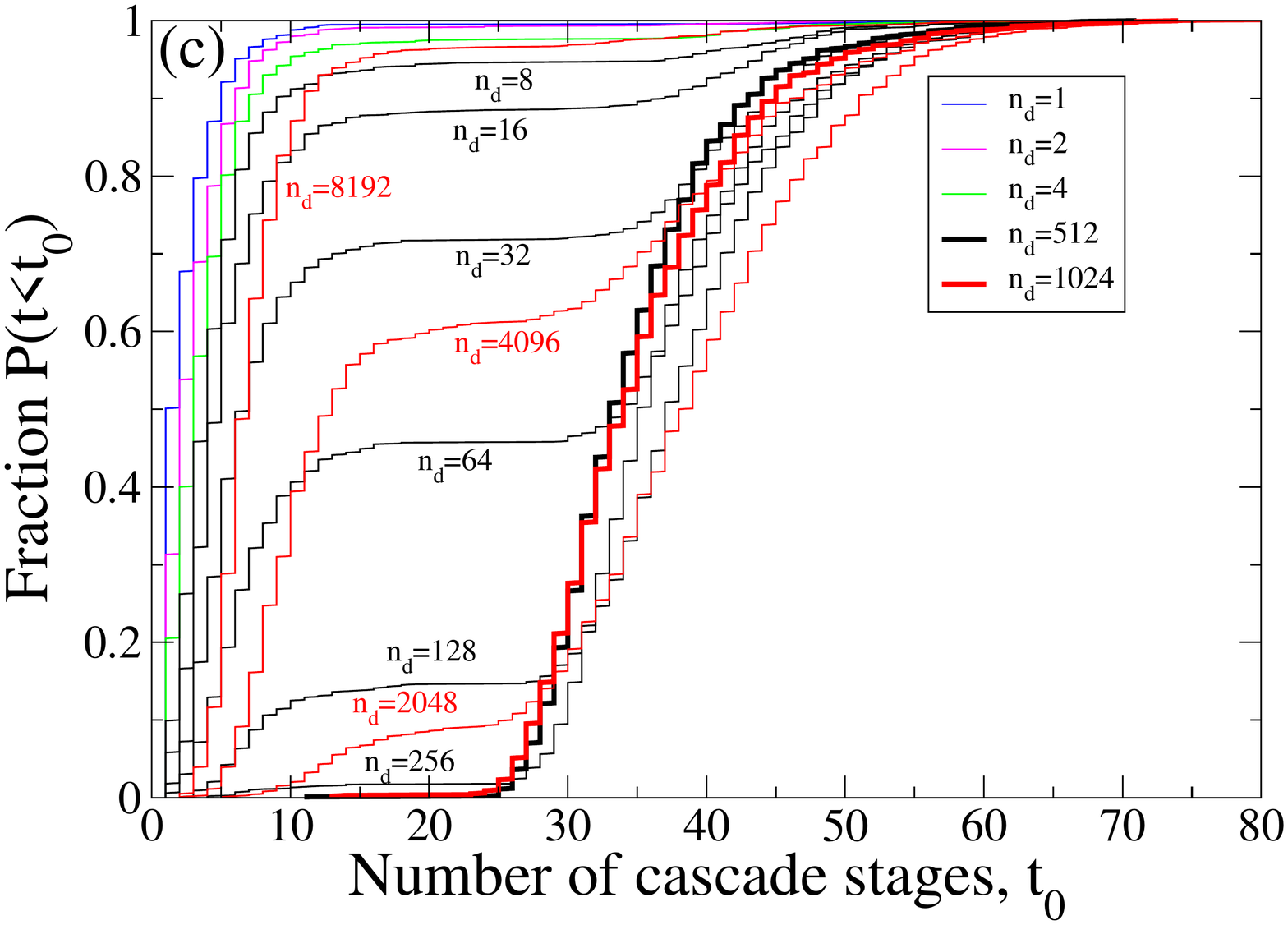}

}

\caption{
 \footnotesize (a) Cumulative distribution of $N_f/N$, the fraction of nodes not overloaded at the end of the cascades
 for different attack size $n_d$,  for $N=20000$, $\langle k\rangle=8$,  and $\alpha=2$. (b) Cumulative distribution for the fraction of nodes in the giant component, $N_c/N$. (c) Cumulative distribution for the number of stages in the cascades $N_t$ for the same set of parameters. The distributions are produced from 1000 independent realizations of the network with a randomly selected central node. The lines corresponding to large $n_d$ for which the fraction of large cascades starts to decrease with increasing $n_d$ are shown in red.
}
\label{fig2}
\end{figure}

We find that for a given value of the parameters $\alpha$, $n_d$, $\langle k\rangle$ and $N$,
some attacks result in the eventual overload of a large fraction of the
nodes; in those cases the giant component of the functional nodes
disintegrates and constitutes a small fraction of functional nodes at
the end of the cascade (Fig.~\ref{fig1}b). Other attacks do not lead to
large cascading failures (Fig.~\ref{fig1}c). The number of stages in
those cascades is small and almost the entire graph and its giant
component survive at the end of them. As a consequence,
there is a wide region in the parameter space in which the
distribution of the cascade lengths,
the distribution of the number of functional nodes at the end of the cascade, and the
distribution of the size of the giant component at the end of the cascade are bimodal, with two peaks for
large and small cascades and nothing in between. In terms of the
cumulative distribution, the bimodality manifests itself by a large
plateau separating the high-slope regions corresponding to small and large cascades
(Fig.~\ref{fig2}).  One can see that the fraction of large cascades, 
indicated by the height of the plateau, increases monotonically with $n_d$ while $n_d$ constitutes a small fraction of $N$. In the particular case studied in Fig.~\ref{fig2} ( $N=20,000$, $\langle k\rangle=8$, $\alpha=2.0$), we observe a monotonic increase of the fraction of large cascades for $n_d<512$. All these 
large cascades result in  65\% of the nodes surviving, almost independently of
$n_d$. The size of the giant component at the end of the cascades constitutes
approximately 20\% of the total number of nodes in the system.
The existence of the plateau suggests that once a cascade reaches a certain number of
failed nodes, it propagates further with probability 100\%. This number
can be estimated from the cumulative distribution of the number of survived nodes,
as the $x$ coordinate of the right end of the plateau which for this system size $N=20,000$,
$\langle k\rangle=8$, and tolerance $\alpha=2$ fluctuates around 19,900. This means that if 100 nodes out
of 20,000 are overloaded the cascade propagates until it destroys the
giant component.

As we can see, the size of the cascade can
dramatically fluctuate for the circular attack of a given size; thus, 
it is impossible to define a critical attack size $n_{dc}$ as such
that for $n_d<n_{dc}$ the cascade is small while for $n_d>n_{dc}$ the
cascade is large. It is more reasonable to define  $P_\ell$,  
the probability of having a large cascade, which is a function of $n_d$,  and then define
 $n_{dc}$  as the value for which 50\% of the cascades are small and 50\%
are large, i.e.  where $P_\ell=0.5$. This condition is satisfied for
$n_d\approx 60$ on Fig.~\ref{fig2}. It should be noted that for larger values of
the tolerance $\alpha$ the plateau in the cumulative distribution may not be as well
defined as in Fig.~\ref{fig2}. In this case $P_\ell$ can be still
defined as the value at which the slope of the cumulative distribution
reaches its minimum (or the minimum in the probability density).

\begin{figure}[h!]
\centering
{
  \includegraphics[width=.9\textwidth]{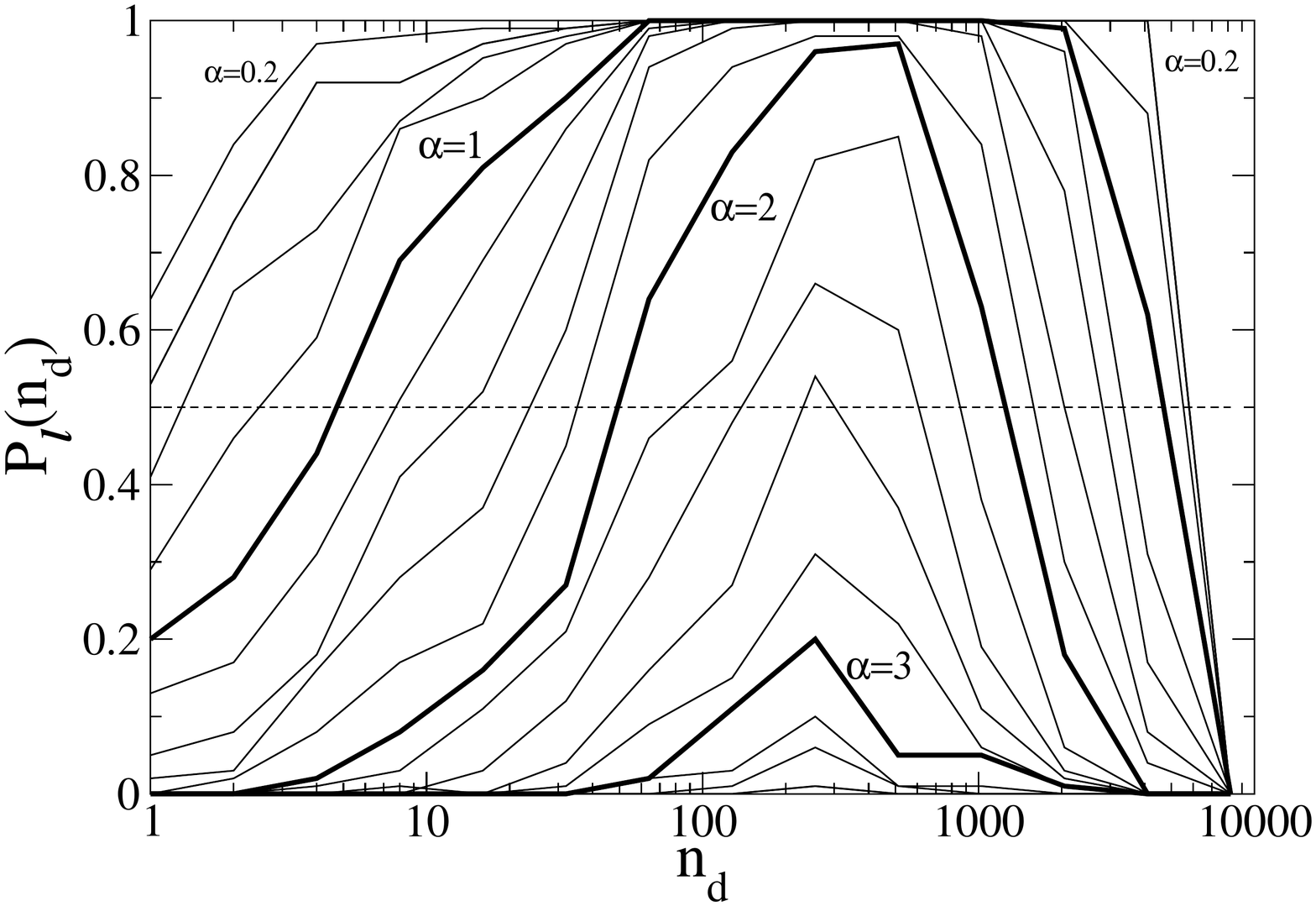}}

\caption{
 \footnotesize The fraction of large cascades, $P_\ell$, as function of the attack size $n_d$ for different tolerances $\alpha$, for $N=10000$ and $\langle k\rangle=8$. The lines corresponding to large $\alpha$ exhibit a maximum at the most efficient attack size $n_{dmax}$ at which $P_\ell< 1$. The horizontal line $P_\ell=0.5$ shows the  crossing which defines the critical attack size $n_d=n_{dc}$.  
For small $\alpha$, $P_\ell$ reaches unity at a certain $n_d>n_{dc}$, and stays at this value until $n_d$ becomes comparable with the system size $N$.   
}
\label{fig3}
\end{figure}

Another question we want to explore is how the probability  $P_\ell$  of large cascades  depends on $n_d$, the size of the attack (Fig.~\ref{fig3}). For small $n_d$, $P_\ell$
increases with $n_d$ with a rate which decreases with $\alpha$.  For
small tolerance $\alpha \leq 2$,  $P_\ell$ reaches 100\% (for the case of $\alpha=1$, it does it at $n_d \approx 80$   in our example of $N=10000$ and 
$\langle k\rangle=8$). After this, $P_\ell$ stays at 100\% until it
starts to decrease for very large values of $n_d$. For larger vales of the tolerance $\alpha$,
 $P_\ell$ increases with $n_d$ until it reaches a maximum, $P_{\ell max}$
at the ``most efficient'' attack size $n_{dmax}$, a quantity which is practically
independent of $\alpha$. As $n_d$ exceed $n_{dmax}$ $P_{\ell}$ starts to decrease to zero
again. For $N=10000$ and $\langle k\rangle=8$, $n_{dmax}=270\pm 20$. As
$\alpha$ increases, $P_{\ell max}$  decreases monotonically. 

\begin{figure}[h!]
\centering
{
  \includegraphics[width=.9\textwidth]{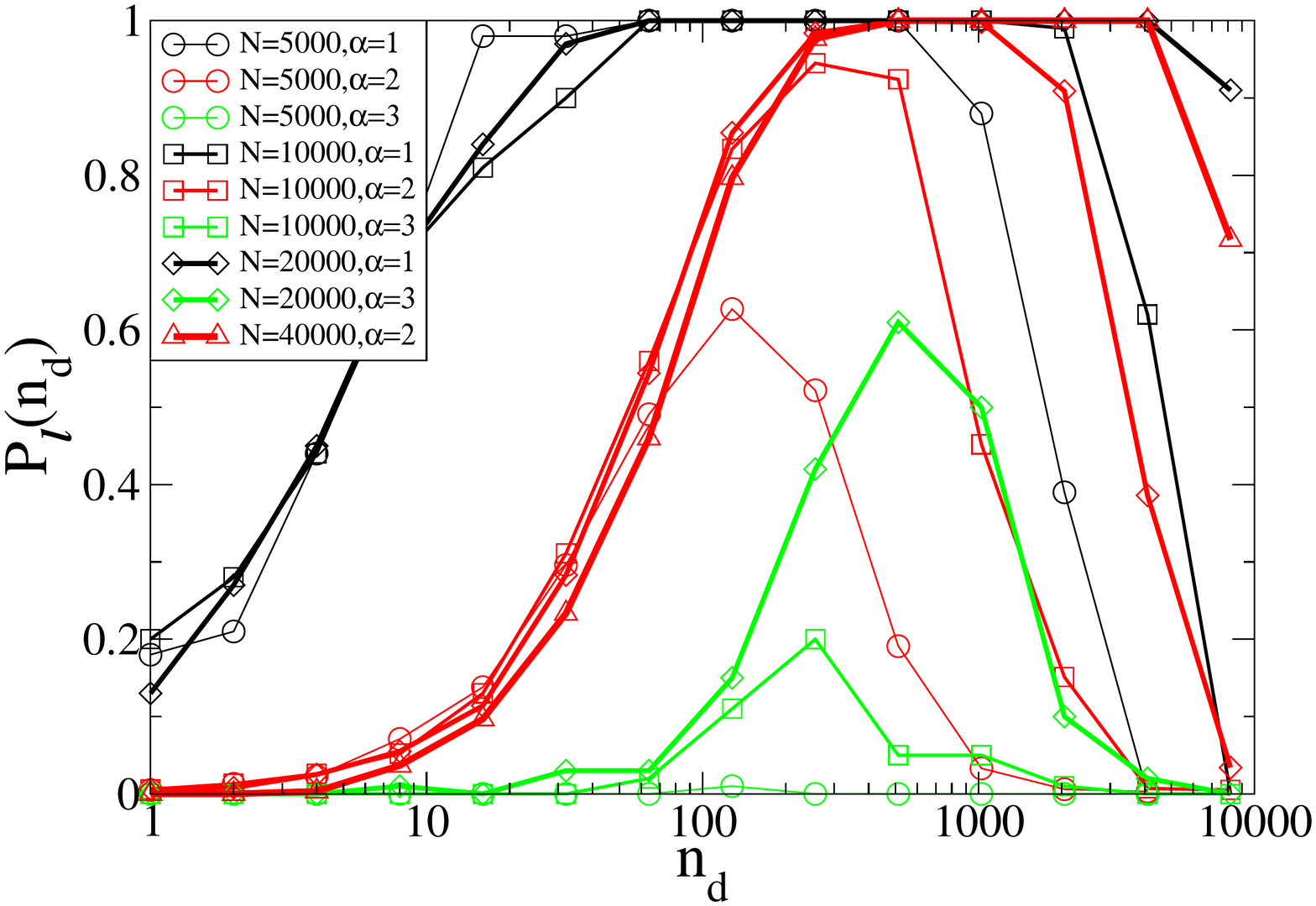}
}
\caption{
 \footnotesize The percent of large cascades as function of the attack size $n_d$ for different tolerances $\alpha$ and different system sizes $N$. Different values of $\alpha$ are indicated by different colors, while different $N$ are indicated by different symbols. One can conjecture that even for $\alpha=3$ (green lines ) for larger $N>20000$, $P_{\ell}$ would reach unity. Also one can see, on the left side of the curves  that for large $N$,  $P_\ell$ becomes a slowly varying function of $N$, which might converge to a fixed limit when $N\to\infty$.  
}
\label{fig4}
\end{figure}

Figure~\ref{fig4} shows that as $n_d$ approaches $n_{dmax}$ the system becomes more vulnerable for
larger system sizes, since for any $n_d$ and $\alpha$, $P_\ell$
increases with $N$, the size of the system.  This is particularly clear in the case of attacks of sizes $ n_d > n_{dmax}$.  The explanation of this phenomenon is the
following. Once $n_d$ becomes comparable with $N$, the amount of
origins and destinations for the traffic becomes $N-n_d$ and the
typical betweenness of nodes becomes roughly proportional to $(N-n_d)^2$ instead of
$N^2$;  accordingly,  when the ratio of these quantities  $(1-n_d/N)^2$
becomes sufficiently smaller than unity,  not a single node will be
overloaded  and no cascades will be observed. From the point of view of real traffic this corresponds to the situation when the area destroyed in the attack is comparable to the size of the entire city. In this case the drivers inside the destroyed area cannot drive, while those drives whose destinations are inside this area cancel their trip. This takes a lot of drivers from the roads and the overall traffic becomes less intense.

\begin{figure}[h!]
\centering
{
  \includegraphics[width=.9\textwidth]{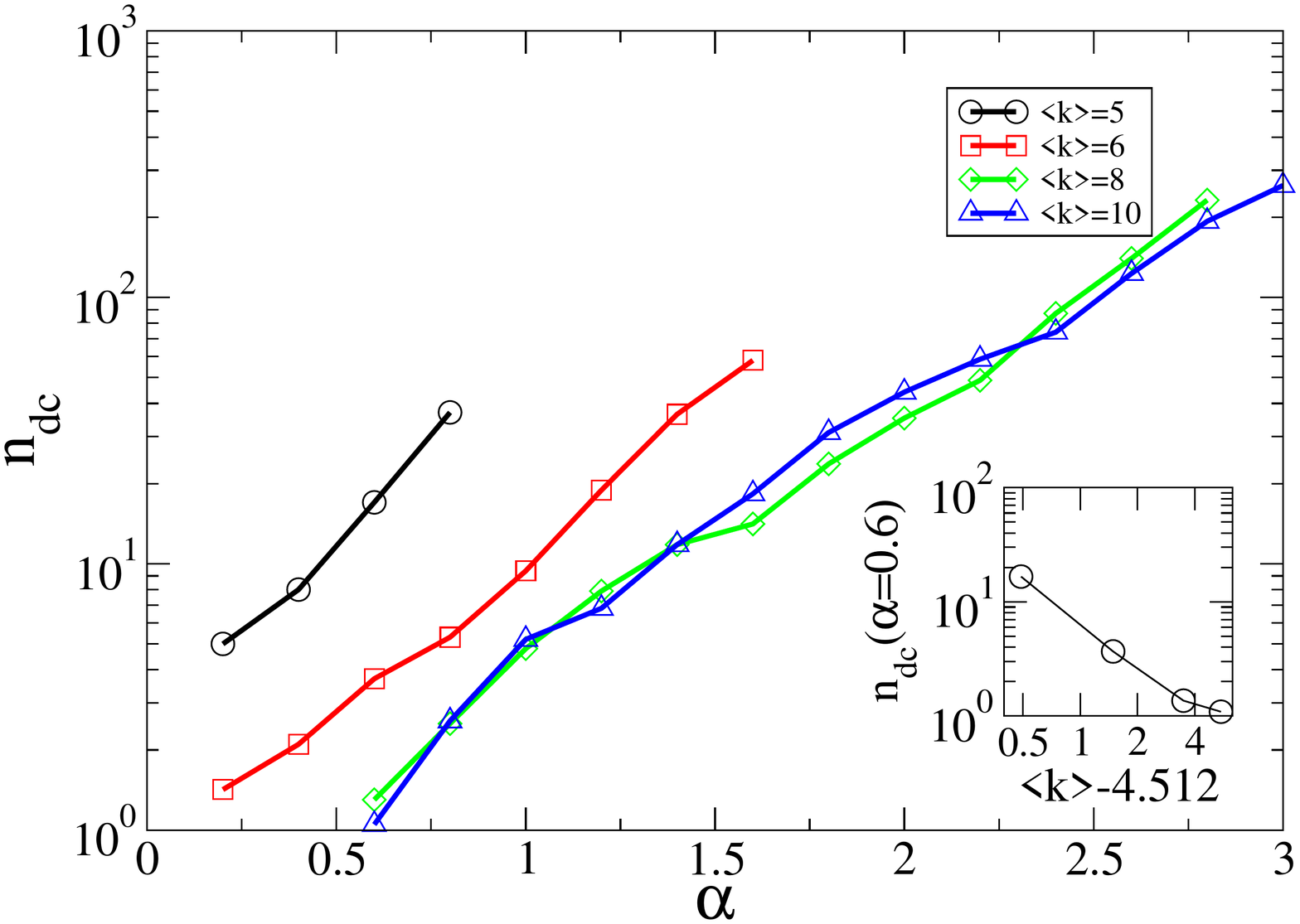}
  }
\caption{
 \footnotesize The behavior of $n_{dc}$ as function of $\alpha$ for different values of  $\langle k\rangle$. Inset: the dependence of $n_{dc}$ as function of $\langle k\rangle -4.512$ for $\alpha=0.6$ as an example of the power law behavior of the size of the 50\% risk attack as function of the distance to the percolation critical point. The slope is approximately equal to 1.2.
}
\label{fig5}
\end{figure}

Indeed Fig.~\ref{fig4} shows  that for smaller $N$ both $P_{\ell max}$
and $n_{dmax}$ become smaller. Accordingly, one can hypothesize that for
$N\to\infty$, $P_\ell$ will become almost independent of the system size and
will increase approximately linearly  with
$\ln[n_d/f(\alpha,\langle k\rangle)]$,  where
$f(\alpha,\langle k\rangle)$ is an increasing function of $\alpha$ and
a decreasing function of $\langle k\rangle$. By definition, $n_{dc}$ is
the value at which $P_\ell=0.5$  and, thus, 
$\ln(n_{dc}) ={\rm const} +f(\alpha,\langle k\rangle)$.  To verify this
relation,  and to find the properties of the function
$f(\alpha,\langle k\rangle)$ we plot $\ln(n_{dc})$ as function of
$\alpha$ for different values of $\langle k\rangle$ for large enough sizes $N$ of the system 
 so that  the dependence on $N$ can be neglected. This is illustrated in 
Fig.~\ref{fig5}. One can see an approximate linear behavior:
$f(\alpha,\langle k\rangle )\approx a(\langle
k\rangle)\alpha+b(\langle k\rangle)$, where $a(k)$ and $b(k)$ are
functions of $k$. While the dependence $b(k)$ is weak, $a(k)$ diverges
when $k\to k_c=4.512$ as a power law, indicating that it is a function of
a correlation length, $\xi$, of the percolation phase transition~\cite{Stauffer}, which
diverges for $k\to k_c$. This can be explained qualitatively by
considering the shortest paths which cross the initial  circle of attack 
of radius $R$ along a certain direction. The number of such independent
paths is $R/\xi$. After the attack, all these paths will be rerouted  through a
node on the perimeter of the circle, and hence its load will increase
by factor $R/\xi$. So for this node to be
overloaded, and the damage done by the initial attack start to spread, $R$ must
satisfy a condition $R>(1+\alpha)\xi$. Thus $n_{dc}$, the size of a critical attack,  should be a
monotonic function of the product of $\alpha$ and
$\xi$. Quantitatively, however, this simple scaling equation is incorrect and requires improvements which we leave out for future studies. From Fig.~\ref{fig5} one can conclude that for the same
tolerance $\alpha$ the system becomes more vulnerable for larger
$\langle k\rangle$ since $n_{dc}$ is a decreasing function of
$\langle k\rangle$.

\subsection{Localized vs. Random Attacks}
In random networks not embedded in space~\cite{Kornbluth2018} the
critical fraction of nodes $1-p=n_{dc}/N$  increases linearly with
$\alpha$ for small values of the tolerance,  and practically does not depend on $N$. Also, it has been found there that  random attacks  are
more efficient than localized ones~\cite{Kornbluth2018}.

\begin{figure}[h!]
\centering
{
  \includegraphics[width=.9\textwidth]{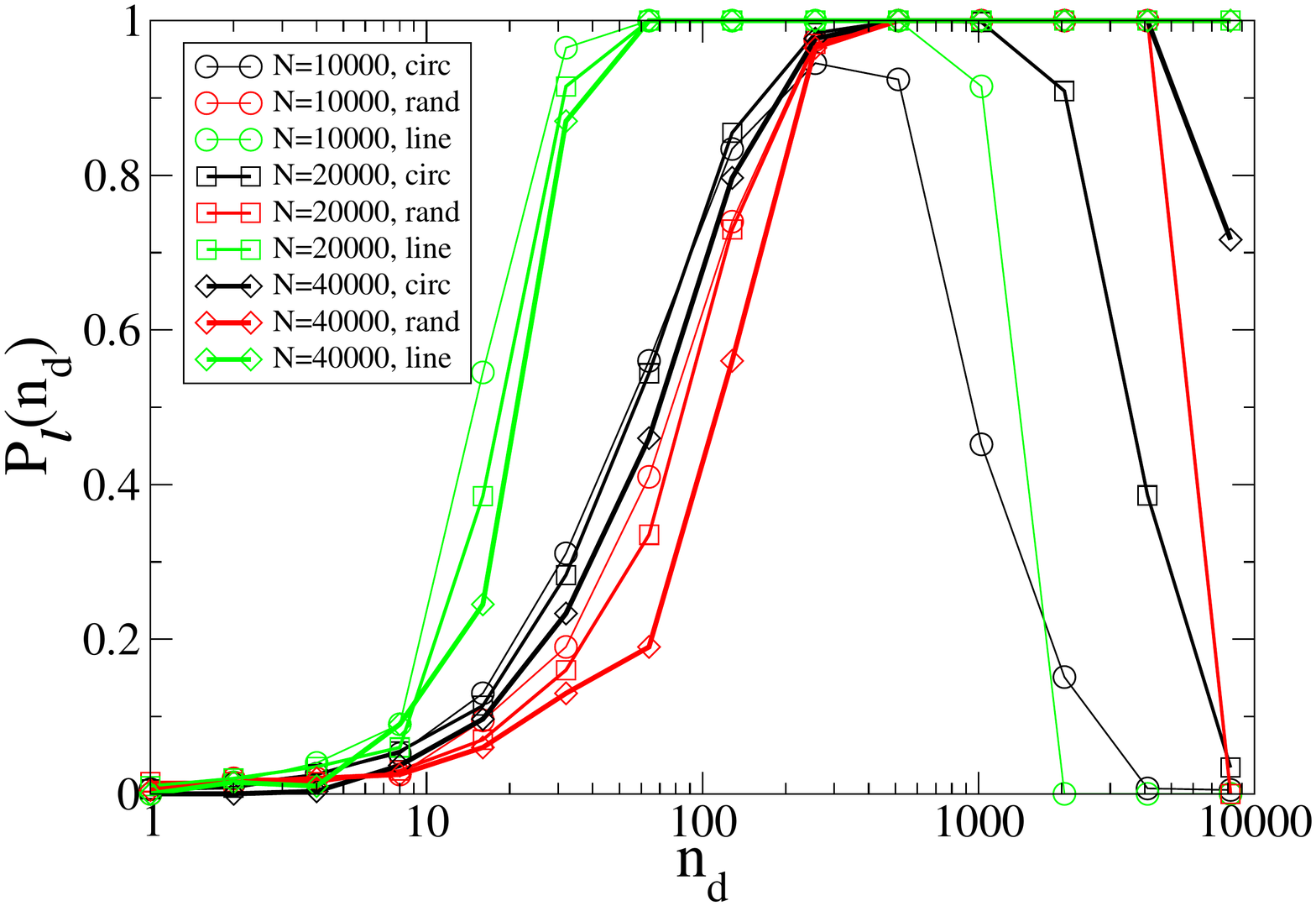}
  }
  \caption{
    \footnotesize Comparison of $P_\ell$ as function of $n_d$ for linear, circular and random attack for different system size $N$,  for $\langle k\rangle=8$ and $\alpha=2$.
}
\label{fig6}
\end{figure}

Here we will examine if these facts are also true  for the Motter and Lai model
on a two-dimensional geometric graph.  Figure~\ref{fig6} compares
the fraction of large cascades $P_\ell$ as function of the size of the initial attack $n_d$ for $\alpha=2$, $\langle k\rangle=8$ and
$N=10000,20000$ and $40000$.  We can see that in the left region of the curves  (where  the vulnerability still increases with the size of the attack)   for
increasing system sizes, $P_\ell$ decreases with $N$ for fixed
$\alpha$ and $n_d$,  i.e. the larger networks are less vulnerable than
smaller ones; and this is true for all three  types of attacks, not only for the circular ones. Therefore, $n_{dc}$ increases with increasing $N$. Still,
$1-p_c=n_{dc}/N$ the fractional size of the attack, decreases when $N$ increases, similarly to the case of random networks.  There  the $1-p_c$ decreased linearly with $\ln N$  with a very small slope. However, in case of the embedded networks studied here, $1-p_c$  decreases with $N$ approximately  as a power law $N^{-0.75}$ for all types of attacks. 
Also, one can see that a random attack is less efficient (or damaging) than a circular one
and that the circular attack is, in turn,  less efficient than the linear one.
The fact that a linear attack is  more efficient than a
circular one is easy to explain. The idea is that the overload
happens, as discussed,  when the linear size $R$ of the attack  is greater than
$(1+\alpha)\xi$, and  for a linear attack this requires fewer attacked nodes $n_d$ than for
the circular attack of the same linear size.  Random attacks were more efficient in terms of destroying a network than localized attacks 
in  random networks not embedded in space;  this  has been explained by taking into
account that random attack eliminated  more links than localized
attacks. Why this is not the case in the networks under study here is not clear, However a similar effect is observed in interdependent networks embedded in space~\cite{Berezin2015,Li2012,Vankin2017}.  This is not a trivial observation, and requires further study.

\subsection{Temporal and Spatial Propagation of the Cascade}

\begin{figure}[h!]
\centering
{
  \includegraphics[width=.49\textwidth]{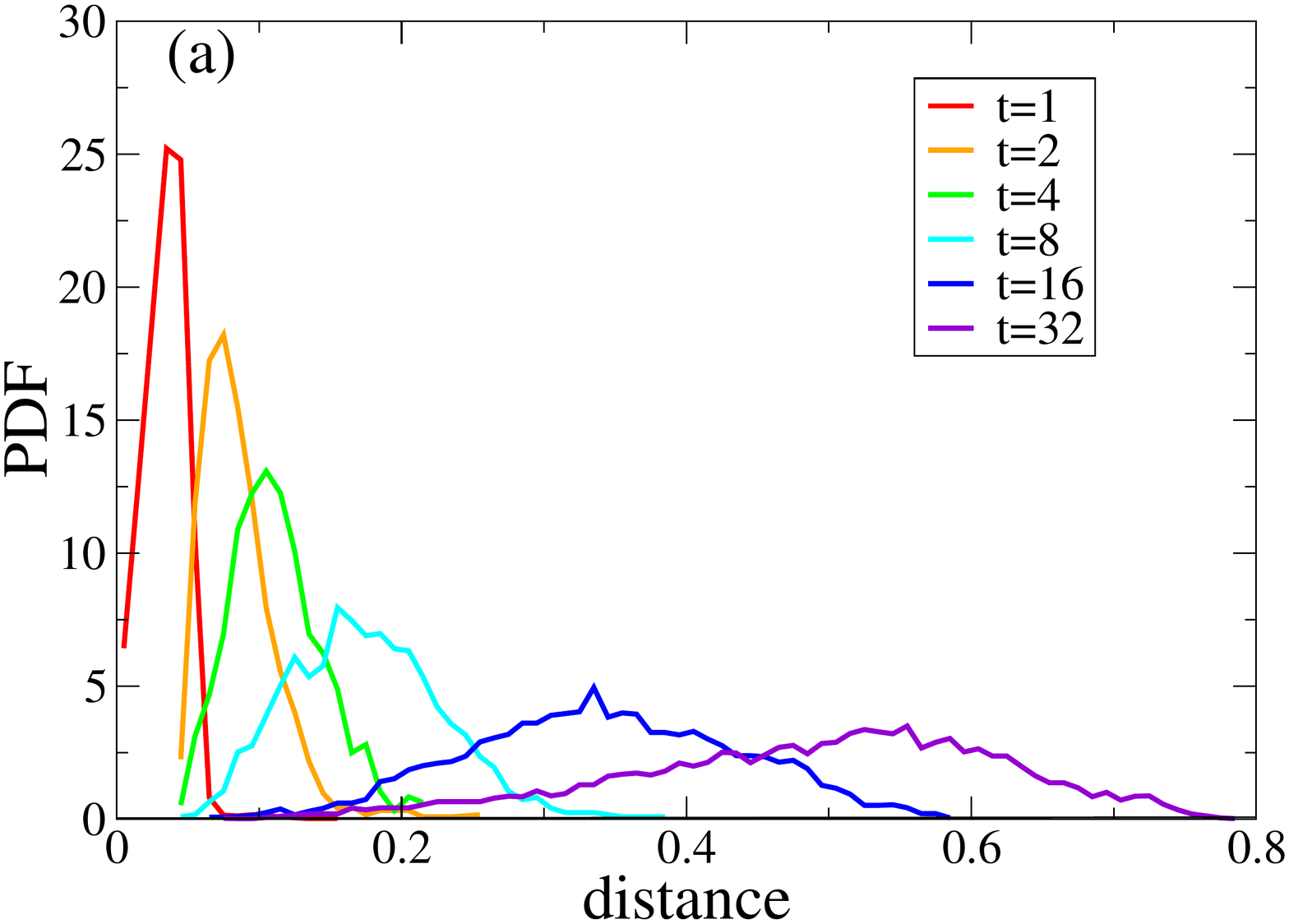}
  \includegraphics[width=.49\textwidth]{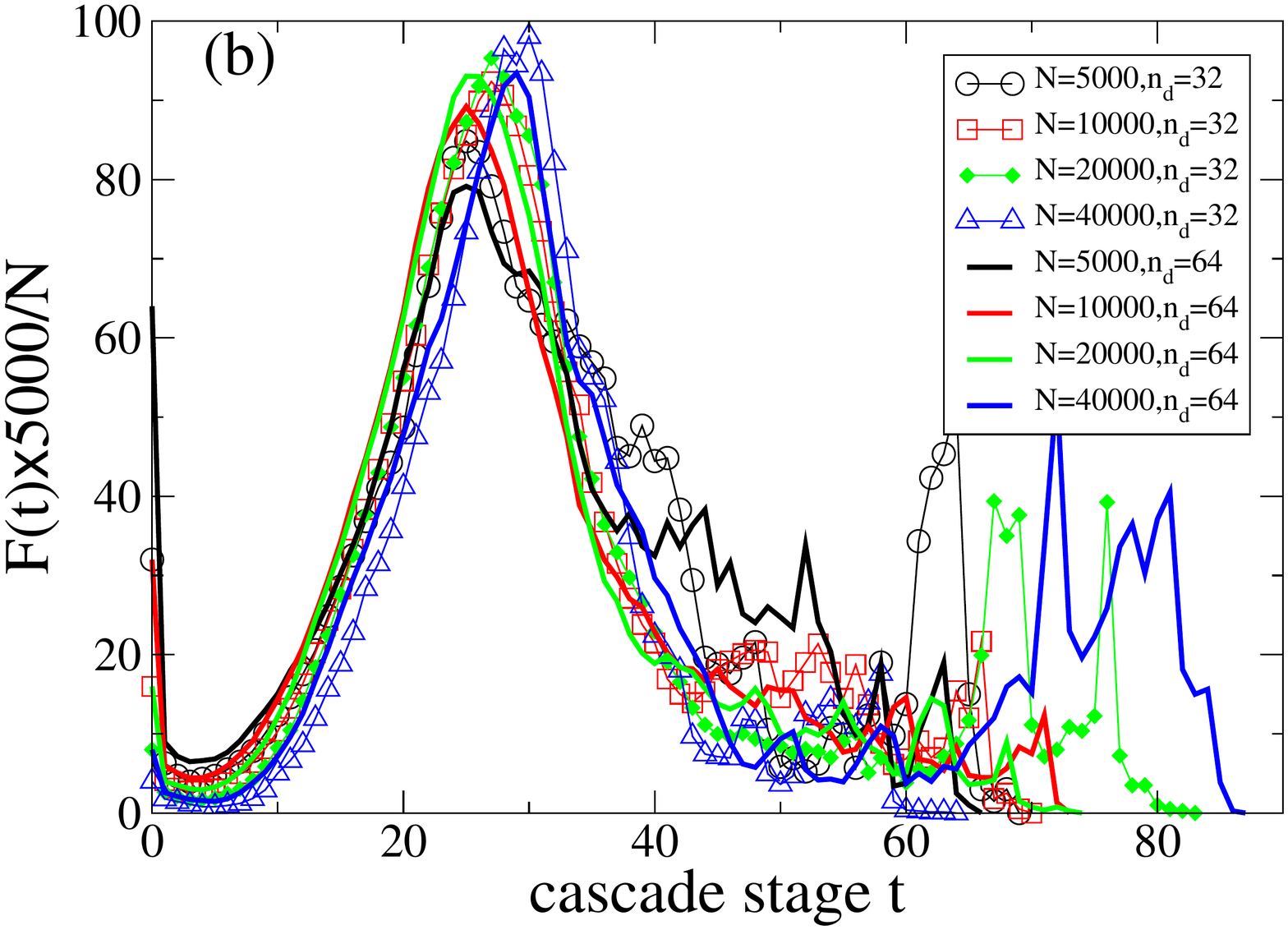}
  \includegraphics[width=.49\textwidth]{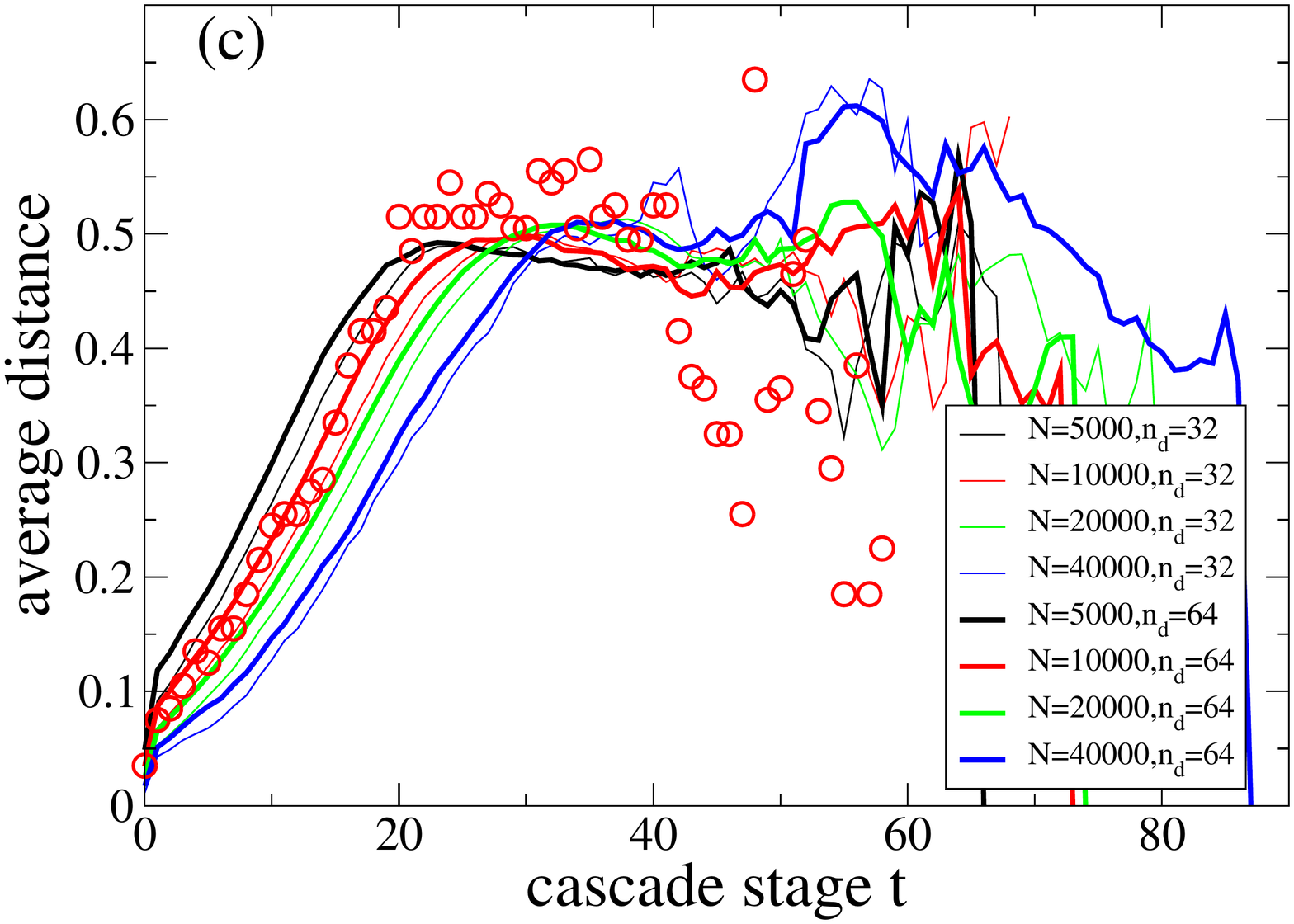}
  \includegraphics[width=.49\textwidth]{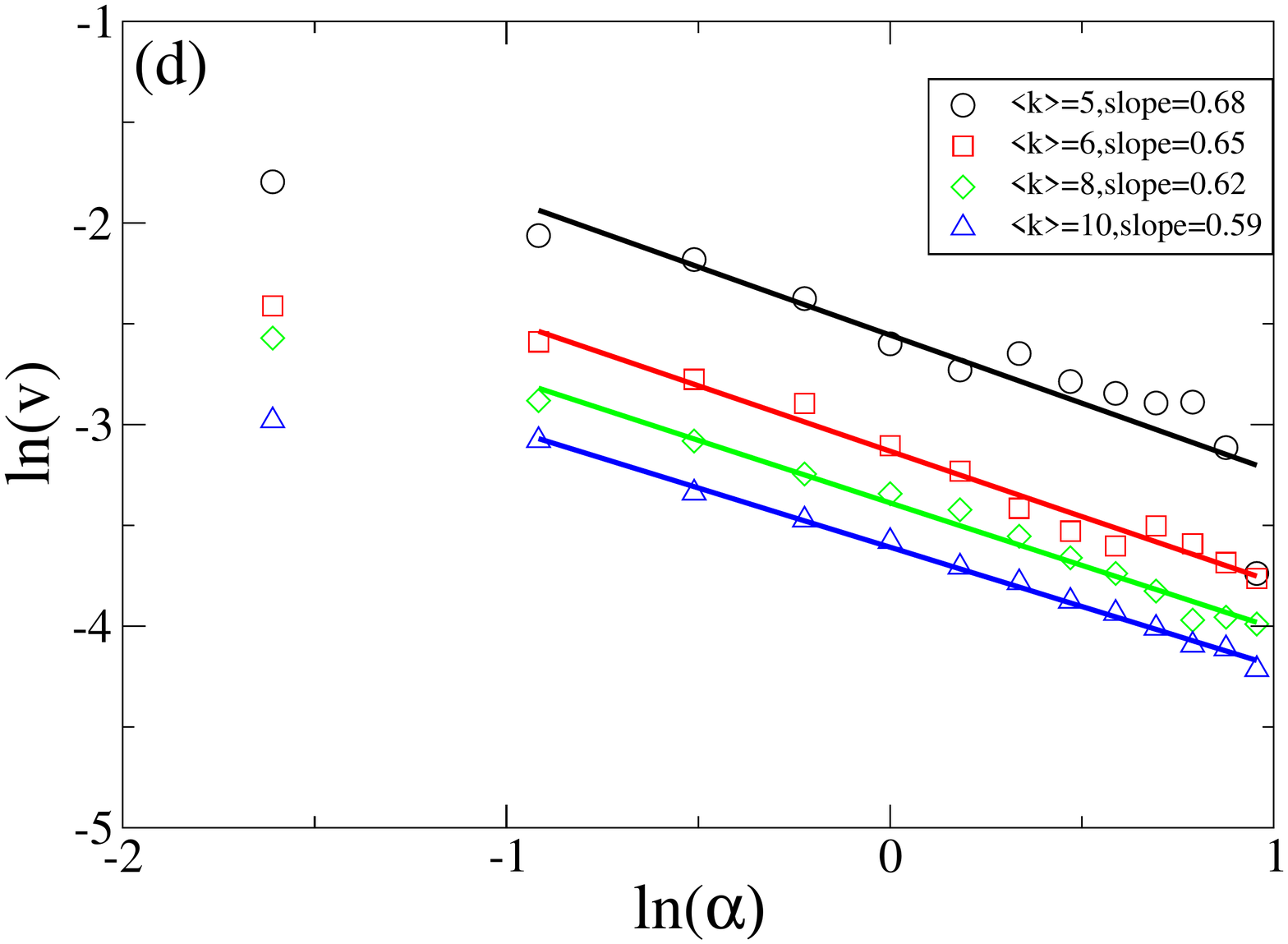}
  }
\caption{
  \footnotesize Propagation of the overload cascades around the circular attack. (a) the PDF of the distances of the overloaded nodes to the central node of the attack, at stages $t$ of the cascade,   for $n_d=64$,  $N=10000$, $\alpha=2.0$ and $\langle k\rangle=8$. (b) The number $F(t)$ of nodes overloaded at stage $t$ of the cascade for different system sizes $N$ and for different sizes of the attack ($n_d$). Note that each curve is scaled to the size of the corresponding system. (c) The dependence of the average distance from the central node of the nodes overloaded at stage $t$ of the cascade  as function of $t$. The circles indicate the most probable distance of the overloaded nodes from the central node at stage $t$, i.e. the position of the peak of the PDFs in panel (a). In (b) and (c) $\langle k\rangle =8$ and $\alpha=2$. (d) Dependence of the velocity of the cascade, (defined as the slope of initial linear part of the graph in panel(c)), with the  tolerance $\alpha$, for different values of $\langle k\rangle$.  $N=10000$ and $n_d=64$.  
}
\label{fig7}
\end{figure}

In a localized attack on a random network not embedded in space
~\cite{Kornbluth2018}, the nodes which are closest to the initial attack have a 
smaller probability of being overloaded during the first stages of the
attack compared with nodes far from the attacked region, while another work~\cite{Havlin} has demonstrated that for the
networks on a square lattice immediately after the initial attack the
overloaded nodes are concentrated near the perimeter of the attack,
while at later stages they spread around the area of the initial
attack as concentric circles. To examine which of these two scenarios
is correct for circular attacks on a random geometric graph, we
compute histograms of the distances from the central node of the attack of the overloaded nodes at the different stages 
$t$ of the cascade [Fig.~\ref{fig7}(a)]
for $N=10000$, $\alpha=2$ and $n_d=64$. The distance of a node from
the central node is computed as the length of the shortest path
connecting these two nodes. Because in our model all the nodes are
located in the unit square, these distance are distributed between 0
and 1. Also, the smallest bin of the distribution in all cases is depleted, because the nodes deleted during the initial attack are all 
concentrated near the central node. For $t=1$ the distribution
has a sharp peak close to the origin. As $t$ increases, the
peak widens and shifts towards the larger distances.

Whether the cascade will spread over the entire system or stop is decided
during its first few stages, when the number of failed nodes at each stage, $F(t)$,
slightly decreases with $t$ [Fig.~ \ref{fig7}(b)]. After this, if the
cascade survives, $F(t)$ starts to grow faster than linearly. $F(t)$  presents a
characteristic maximum near the middle stages of the cascade. The
number of nodes in this maximum grows as $N$, the same as 
observed for the $\zeta$ model on lattice graphs~\cite{Havlin}, while  the
width of the peak stays constant with the size of the system;  these two facts indicate that  
the total fraction of the overloaded nodes at the end of the
cascade is independent of the system size. 

The average distance of the nodes overloaded at stage $t$ increases linearly with $t$ at the beginning,  while the size of the overloaded area is much smaller than the  size of the system [Fig.~ \ref{fig7}(c)]. For small $t$ many cascades propagate slowly, being on a brink of extinction, especially for $n_d<n_{dc}$. After this initial latent period of the cascade, after $F(t)$ reaches its minimum and before it reaches its maximum (which happens when the size of the overloaded area becomes comparable to the system size), the velocity of the cascade propagation is approximately constant. The average velocity in this regime is a slightly decreasing function of the size of the system $N$. This is due to the fact that the linear size of the system in our simulations does not depend on $N$, and the stage $t_{\rm max}$ when $F(t)$ reached its maximum is practically independent on $N$, while the total number of nodes overloaded by this time is proportional to $N$. Accordingly, the overloaded nodes are distributed in the same number of growing fingering shapes corresponding to different stages of the cascade, which for larger $N$ contain larger number of nodes. Thus one can expect that the velocity of the cascade spatial propagation would depend very mildly on $N$. In contrast with that, one can expect that the velocity $v$ of propagation of the cascade should decrease when the tolerance $\alpha$ increases. Indeed, we found (Fig.~\ref{fig7}(d)), that $v \approx v_0(\langle k\rangle)\alpha^{-0.6}$ for the networks with different average degrees.  The prefactor $v_0(\langle k\rangle)$ diverges as $(\langle k\rangle - \langle k\rangle_c)^{-0.43}$ as a function of the correlation length.   

\begin{figure}[h!]
\centering
{
  \includegraphics[width=.9\textwidth]{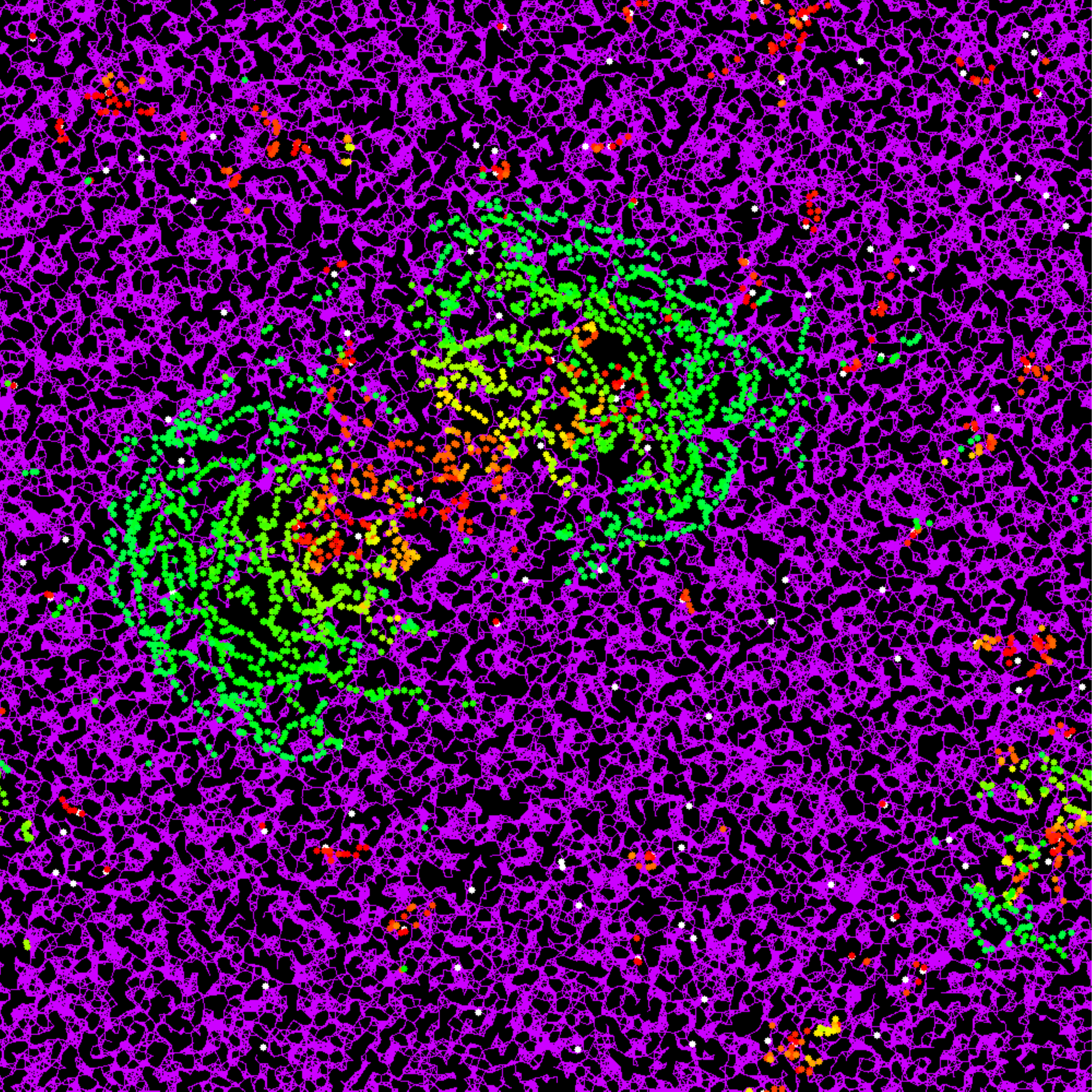}
  }
\caption{
  \footnotesize A snapshot of a system of $N=40000$ nodes at stage $t=19$ of the cascade of  overloads, following  a random initial attack with $n_d=128$.   $\alpha=2.0$ and $\langle k\rangle=8$.   
}
\label{fig8}
\end{figure}

Figure~\ref{fig1} shows
that the nodes deleted at each stage of the cascade follow some shortest
``roads'' avoiding the previously overloaded area. Thus, the fastest
growth occurs near the tips of the previously overloaded area and, hence, the
growth of the overloaded area somewhat resembles viscous fingering~\cite{viscous}.
 It is not
surprising that for the linear and the circular attacks this growth
mechanism is similar. What is unexpected is that for the random
attack, after a first few stages when the overloads are randomly
distributed over the entire system, these overloads become concentrated
near fixed centers, after which the growth of the overloaded area
starts to resemble the viscous fingering of the circular 
attacks around these centers (Fig.~\ref{fig8}). Obviously, the type of instability in the Motter and Lai model differs from the Saffman-Taylor instability~\cite{viscous}, but it is definitely not the circular growth observed in~\cite{Havlin} for a lattice model and resembles the equipotential lines of a dipole.
 
\section{Conclusion}
We investigate the Motter and Lai model on a two-dimensional random geometric graph with average degree $\langle k\rangle $ for the case of circular,
linear and random initial attacks. We find that the distribution of the sizes of the cascades is bimodal, consisting of
very small cascades and very large cascades which destroy a finite fraction of the entire system,
and nothing in between. We define the critical size of the attack $n_{dc}$ as
the number of initially deleted nodes at which the probability of large cascades $P_\ell$ is 50\%. We
find the following qualitative properties of the system.

(i) The random attacks are less efficient than circular attacks which are, in turn, less efficient than linear attacks.

(ii) The larger systems are less vulnerable than smaller systems for the same $\alpha$ and $\langle k\rangle$ if $n_d\ll N$.

(iii) In a marked difference with the networks not embedded in space, the critical size of the attack $n_{dc}$ is not proportional to $N$ as it is for the not embedded networks, but only mildly increases with $N$.  

(iv) The densely connected graphs with large average degree $\langle k\rangle$ are more vulnerable than the graphs with smaller average degree.

(v) The critical size of the attack, $n_{dc}$, increases exponentially with the tolerance $\alpha$. 

(vi) For large enough $\alpha$, there is a most efficient attack size at which $P_\ell$ reaches a maximum. The height of this maximum, $P_{\ell max}$,  increases with the system size $N$, until it reaches 100\%.

(vii) At the beginning of a localized attack the overloads are concentrated near the area initially attacked.

(viii) At later stages of the cascade, the growth of the overloaded region is not circular but resembles viscous fingering around a few centers for the case of a random attack, or around the central node in the case of a localized attack (both for circular or linear attacks).

(ix) The spatial speed of the propagation of the cascade, for the case of circular and linear attacks, is practically independent of $N$, but it is inverse proportional to a power of the tolerance $\alpha$, and proportional to a power of a correlation length.

(x) Near the critical attack size $n_{dc}$, the cascades have an initial latent period during which the number of overloaded links at each step fluctuates near zero and the cascade may spontaneously stop.

All these qualitative relations require further detailed investigation in order to determine accurate scaling laws behind them. 

\section{Acknowledgements}

This research is supported by the Defense Threat Reduction Agency grant HDTRA1-19-1-0016 and by Binational Science Foundation Grant No. 2020255. We also acknowledge the partial support of this research through the Dr. Bernard W. Gamson Computational Science Center at Yeshiva College. We thank A. Bashan, D. Vaknin and S. Havlin for useful comments.

\bigskip
\newpage

\bigskip
\newpage

\bigskip
\newpage
\bigskip
\newpage
\bigskip
\newpage
\bigskip
\newpage
\bigskip
\newpage
\bigskip
\newpage
\bigskip
\end{document}